\documentclass[prl,twocolumn,showpacs,superscriptaddress,floatfix]{revtex4}
\usepackage{graphicx,amssymb,amsmath}

\newcommand{\crd}{d^\dagger}
\newcommand{\annd}{d^{\phantom\dagger}}

\newcommand{\crb}{b^\dagger}
\newcommand{\annb}{b^{\phantom\dagger}}

\newcommand{\crs}{s^\dagger}

\newcommand{\pal}{\partial_l}

\newcommand{\crP}{P^\dagger}
\newcommand{\annP}{P^{\phantom\dagger}}

\begin{document}
\title{Finite-size scaling exponents in the interacting boson model}

\author{S\'ebastien Dusuel}
\affiliation{Institut f\"ur Theoretische Physik, Universit\"at zu
K\"oln, Z\"ulpicher Str. 77, 50937 K\"oln, Germany}

\author{Julien Vidal}
\affiliation{Laboratoire de Physique Th\'eorique de la Mati\`ere Condens\'ee, CNRS UMR 7600,
Universit\'e Pierre et Marie Curie, 4 Place Jussieu, 75252 Paris Cedex 05, France}

\author{Jos\'e M. Arias}
\affiliation{Departamento de F\'{\i}sica At\'omica, Molecular y
Nuclear, Facultad de F\'{\i}sica, Universidad de Sevilla,
Apartado~1065, 41080 Sevilla, Spain}

\author{Jorge Dukelsky}
\affiliation{Instituto de Estructura de la Materia, CSIC, Serrano
123, 28006 Madrid, Spain}

\author{Jos\'e Enrique Garc\'{\i}a-Ramos}
\affiliation{Departamento de F\'{\i}sica Aplicada, Universidad de
Huelva, 21071 Huelva, Spain}

\begin{abstract}
We investigate the finite-size scaling exponents for the critical
point at the shape
phase transition from U(5) (spherical) to O(6) (deformed
$\gamma$-unstable) dynamical symmetries of the Interacting Boson
Model, making use of the Holstein-Primakoff boson expansion and
the continuous unitary transformation technique. We compute
exactly the leading order correction to the ground state energy,
the gap, the expectation value of the $d$-boson number in the
ground state and the $E2$ transition probability from the ground
state to the first excited state, and determine the
corresponding finite-size scaling exponents.
\end{abstract}


\maketitle


The interest in the study of quantum phase transitions (QPT) has
kept growing in the last years in different branches of quantum
many-body physics, ranging from macroscopic systems like quantum
magnets, high-$T_\mathrm{c}$ superconductors \cite{Vojta} or
dilute Bose and Fermi gases \cite{Sad} to mesoscopic systems such
as atomic nuclei or molecules \cite{IZ04}. Although, strictly
speaking QPT only occur in macroscopic systems, there is a renewed
interest in studying structural changes in finite-size systems
where precursors of the transition are already observed
\cite{Casten}. The understanding of the modifications on the
characteristics of the QPT induced by finite-size effects is of
crucial importance to extend the concept of phase transitions to
finite systems.

In the present study, we analyze these finite-size corrections in
the Interacting Boson Model (IBM) of nuclei \cite{IBMbook},
but the same technique can be applied to other boson systems,
for instance to the molecular vibron model \cite{Molbook}, or to a
multilevel boson model of Bose-Einstein condensates where similar QPT
take place \cite{Duke1}.

The IBM is a two-level boson model that includes an angular momentum
$L=0$ boson (scalar $s$-boson) and five angular momentum $L=2$
bosons (quadrupole $d$-bosons) separated by an energy gap. The $s$
and $d$-bosons represent $s$ and $d$ wave idealized Cooper nucleon
pairs. The algebraic structure of this model is governed by the
U(6) group and the model has three dynamical symmetries in which
the Hamiltonian, written in terms of the invariant (Casimir)
operators of a nested chain of subgroups of U(6), is analytically
solvable. The dynamical symmetries are named by the first subgroup
in the chain: U(5), SU(3) and O(6). The classical or thermodynamic
limit of the model was investigated by using an intrinsic state
formalism that introduces the shape variables $\beta$ and $\gamma
$ \cite{GK80,DSI80,BM80}. Within this geometric picture the U(5),
SU(3) and O(6) dynamical symmetries correspond to spherical,
axially deformed and deformed $\gamma$-unstable shapes,
respectively. Transition between two of these  dynamical symmetry
limits are described in terms of a Hamiltonian with a control
parameter that mixes the Casimir operators of the two dynamical
symmetries. As a function of the control parameter, the system
crosses smoothly a region of structural changes in the ground
state wave function for finite number $N$ of bosons. In the large
$N$ limit,  the smooth crossover turns into a sharp QPT between
two well defined shape phases
\cite{DSI80,FGD81,Alex89,LMC96,Jo02}. In particular, the
transition from U(5) to O(6) has been intensively studied in
recent years because it has a unique second order QPT
\cite{LMC96,Jo02,ADG03} associated with a triple point in the IBM
parameter space \cite{Jo02}. Furthermore, it was early recognized
that the IBM Hamiltonian along this transition was fully
integrable \cite{Alha}, and exactly solvable \cite{Pan,Duke2}.

Unfortunately, it is difficult to use the exact solution to
 compute finite-size corrections analytically. Thus, we follow
a different route which is based on the Continuous Unitary
Transformations (CUTs) \cite{Wegner94,Glazek93,Glazek94}. Within
this framework, we compute the first correction beyond the
standard Random Phase Approximation (RPA) \cite{Duke84} which
already contains the key ingredients to analyze the critical
point.  As already observed in a similar context,
\cite{Dusuel04_3,Dusuel05_1,Dusuel05_2}, this $1/N$ expansion
becomes, at this order, singular when approaching the critical
region so that one gets nontrivial scaling exponents for the
physical observables (ground state energy, gap, occupation number,
transition rates). In a second step, we take advantage of the
exact solvability of the model to obtain
numerical results for large number of bosons which allows us to check our analytical
predictions. 

Let us consider the U(5)-O(6) transitional Hamiltonian
\begin{equation}
  \label{eq:hamiltonian}
  H=x \, n_d+\frac{1-x}{4(N-1)}\left( \crP_d-\crP_s\right)
  \left(\annP_d-\annP_s\right),
\end{equation}
where $n_d$ $\sum_\mu \crd_\mu \annd_\mu$ (with $\mu=-2,-1,0,1,2$) is the $d$-boson number operator, $\crP_s={\crs}^2$ and
$\crP_d=\sum_\mu (-1)^\mu \crd_\mu \crd_{-\mu}$~; $x$ is the
control parameter that mixes the U(5) linear Casimir operator
 ($x=1$) with the O(6) quadratic Casimir operator ($x=0$). The system
undergoes a QPT at $x_\mathrm{c}=1/2$, between a U(5) (spherical)
phase for $1/2 \leq x \leq 1$ and a O(6) (deformed
$\gamma$-unstable) phase for $0 \leq x \leq 1/2$, when
$N\to\infty$ \cite{LMC96,Jo02,ADG03}.


In the following, we shall restrict our analysis to the spherical
(symmetric) phase which allows to investigate the critical point
more simply than from the deformed phase. The Holstein-Primakoff
boson expansion of one-body boson operators is especially
well-suited to perform a $1/N$ expansion of the boson Hamiltonian
(\ref{eq:hamiltonian}). In the present case, it reads
\cite{Holstein40,Klein91}
\begin{subequations}
  \label{eq:HP}
  \begin{eqnarray}
    \crd_\mu \annd_\nu &=& \crb_\mu \annb_\nu ,\\
    \crs s &=& N-\sum_\mu \crd_\mu \annd_\mu=N-\sum_\mu \crb_\mu \annb_\mu
    =N-n_b,\quad\quad\\
    \crd_\mu s &=& N^{1/2} \crb_\mu (1-n_b/N)^{1/2}
    = (\crs \annd_\mu)^\dagger.
  \end{eqnarray}
\end{subequations}
Keeping terms of order $(1/N)^0$ in the Hamiltonian expressed in
terms of the new $b$'s yields a quadratic Hamiltonian which can be
diagonalized via a Bogoliubov transformation. One then recovers
RPA results \cite{Duke84,Rowe04_1}. At the next order  $(1/N)^1$,
the Hamiltonian is quartic and diagonalizing it clearly requires a
more sophisticated method. To achieve this goal,  we used the CUTs
technique \cite{Wegner94,Glazek93,Glazek94}.  For an introduction
to this method, we refer the reader to
Refs.~\cite{Dusuel04_3,Dusuel05_1,Dusuel05_2} where CUTs were
applied in a similar context. One introduces a running Hamiltonian
\begin{eqnarray}
  \label{eq:running_ham}
  H(l) &=& E_0(l) + \Delta(l) n_b + V(l) :n_b^2: + W(l) \crP_b \annP_b\\
  &&+\Gamma(l)\left(\crP_b + \annP_b \right)
  + \Lambda(l) \left(\crP_b n_b + n_b \annP_b\right),\quad\nonumber
\end{eqnarray}
which is related to the initial Hamiltonian $H(0)$ through a
unitary transformation, namely $H(l)=U^\dagger(l) H(0) U(l)$. This
transformation $U$ is chosen such that $H(\infty)$ commutes with
$n_b$. In Eq.~(\ref{eq:running_ham}), $:\mathcal{O}:$ denotes the
normal ordered form of the operator $\mathcal{O}$, and the
notations for the $b$'s are the  same as for the $d$'s. The
evolution of the running Hamiltonian is obtained from the flow
equation $\partial_l H(l)=[\eta(l),H(l)]$, where $\eta(l) =
\pal U^\dagger(l)  U(l)$ is the anti-hermitian generator of
the unitary transformation. For the problem at hand, we consider
the so-called quasi-particle conserving generator
\cite{Knetter00},
\begin{equation}
  \label{eq:generator}
  \eta(l) = \Gamma(l)\left(\crP_b - \annP_b \right)
  + \Lambda(l) \left(\crP_b n_b - n_b \annP_b\right),
\end{equation}
designed to ensure $H(\infty)$ commutes with $n_b$, {\em i.e.}
$\Gamma(\infty)=\Lambda(\infty)=0$.

The flow equations can be solved exactly, order by order in $1/N$,
and the coefficients of the final Hamiltonian are found to be
\begin{eqnarray}
  \label{eq:Einf}
  E_0(\infty) &=& \frac{N(1-x)}{4} + \frac{5}{2} \left[\frac{1}{2}(1-3x)
    +\Xi(x)^{1/2} \right]\\
  &+& \frac{5x (1-x)}{N}\left[\frac{25x-9}{16\Xi(x)}
    -\frac{1}{\Xi(x)^{1/2}}\right],\nonumber\\
  \label{eq:Dinf}
  \Delta(\infty) &=& \Xi(x)^{1/2}+ \frac{x(1-x)}{N}\left[\frac{9x-1}{4\Xi(x)}
    -\frac{2}{\Xi(x)^{1/2}}\right],\quad\\
  \label{eq:Vinf}
  V(\infty) &=& \frac{x^2(1-x)}{4 N \Xi(x)},\\
  \label{eq:Winf}
  W(\infty) &=& \frac{x(1-x)(3x-1)}{8 N \Xi(x)},
\end{eqnarray}
where $\Xi(x)=x(2x-1)$. One can then straightforwardly analyze the
low-energy spectrum.

\begin{figure}[h]
  \centering
  \includegraphics[width=8cm]{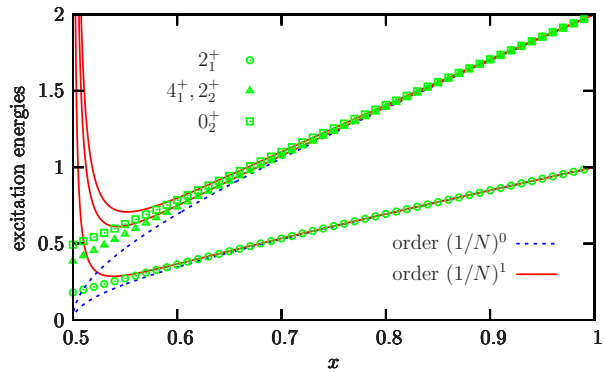}
  \caption{Comparison between analytical results (solid and dotted lines) and the numerical
  results (circle, triangle and square symbols) for the first excitation energies, with $N=40$.}
  \label{fig:exc}
\end{figure}

The ground state of $H(\infty)$ is the state $|0\rangle$ with zero
$b$-bosons, whose energy is $E_0(\infty)$.
The first excited state is five-fold degenerate and corresponds to
one quadrupole boson $\crb_\mu |0\rangle$, whose excitation energy
is $\Delta(\infty)$. These are the five components of the first
$2^+$ excited state.
For the two-boson states, things are a bit more complicated
because of the $W$ term which is not diagonal in the basis of
states $\{ \crb_\mu \crb_\nu |0\rangle\}$ with
$\mu,\nu=-2,-1,0,1,2$. It is however easy to see that $\crP_b
\annP_b$has a nontrivial action  only in the subspace
$\Big\lbrace \crb_2 \crb_{-2}|0\rangle, \crb_1 \crb_{-1}|0\rangle,
\frac{1}{\sqrt{2}}{\crb_0}^2|0\rangle\Big\rbrace$. The
corresponding $3\times 3$ matrix has eigenvalues $0$ (twice) and $10$. One
thus finds that there are 14 degenerate states with excitation
energy $2[\Delta(\infty)+V(\infty)]$, and one $0^+$ state which is given by
$\frac{1}{\sqrt{10}}\crP_b|0\rangle$ with energy
$2[\Delta(\infty)+V(\infty)+5W(\infty)]$. Let us emphasize that
the degeneracy is lifted at order $(1/N)^1$, an effect missed at
the RPA order. The 14 degenerate states are the 9 components of
the first excited $4^+$ state and the 5 components of the second
excited $2^+$ state. These $4_1^+$ and $2_2^+$ states are
degenerate along the whole transition line because of the common O(5)
structure. Note also that, at fixed $N$,  the $0^+_2$ state
degenerates with the $4_1^+$ and $2_2^+$ in the U(5) limit. This
low-energy spectrum is depicted in Fig.~\ref{fig:exc} for $N=40$.
The agreement between numerics and analytical results is pretty
good and has been checked to improve when $N$ gets bigger, as long
as one is sufficiently far away from the critical point. Indeed,
as can be seen in (\ref{eq:Einf}-\ref{eq:Winf}),  the $1/N$ order
corrections diverge at $x=1/2$. This singular behavior already
found in other models \cite{Dusuel04_3,Dusuel05_1,Dusuel05_2} is a
signature of the noninteger scaling exponents \cite{Rowe04_2} that
we discuss below.

\begin{figure}[h]
  \centering
  \includegraphics[width=8cm]{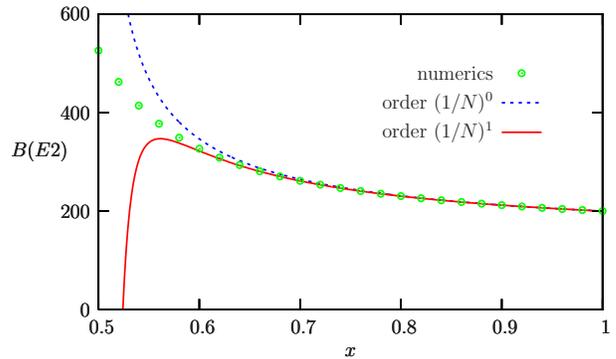}
  \caption{Comparison between analytical results (solid dotted lines) and the numerical
  results (circles) for the $B(E2)$ transition probability, with $N=40$.}
  \label{fig:be2}
\end{figure}

The main strength of the CUTs is to allow the computation of
expectation values of observables as well as transition
amplitudes. Thus, one has to perform the unitary
transformation of the observables in which one is interested. In
the present case, all observables can be deduced from the
knowledge of the flow of the operator $\crb_\mu(l)=U^\dagger(l)
\crb_\mu U(l)$. For example, the average number of $d$-bosons in
the ground-state  of the Hamiltonian $H$ is found as $\langle n_d
\rangle=\langle 0| \sum_\mu \crb_\mu(\infty)\annb_\mu(\infty)
|0\rangle$. This quantity can also be computed using the
Hellmann-Feynman theorem which yields $\langle n_d
\rangle=\frac{\partial}{\partial y} {[(1+y)E_0]}$ with
$y=x/(1-x)$. One then gets
\begin{eqnarray}
  \label{eq:nd}
\langle n_d \rangle&=&
  \frac{5}{2}\left[\frac{3x-1}{2 \Xi(x)^{1/2}}-1\right] \\
  &&+\frac{5x (1-x)^2}{16N}\left[ -\frac{7x}{\Xi(x)^2} +\frac{8}{\Xi(x)^{3/2}}\right].
  \nonumber
\end{eqnarray}
However, this theorem cannot be applied to compute nondiagonal
matrix elements such as transition amplitudes. To illustrate the
power of the CUTs for such a task, we focus on the $B(E2)$
transition probability between the ground state and the first
excited state which is defined as
$B(E2)=5\big|\langle2,0|Q_0^{(2)}|0,0\rangle \big|^2$ in the
standard $|J,M\rangle$ basis, with $Q_0^{(2)}=\crs\annd_0+\crd_0
s$. The flow equations for $\crb_\mu(l)$ can still be exactly
integrated out order by order in $1/N$ and  leads to:
\begin{eqnarray}
   \label{eq:be2}
  B(E2) &=&
  N\frac{5x}{\Xi(x)^{1/2}}\\
  &&+5x^2\left[ -\frac{27x^2-20x+5}{4\Xi(x)^2}
  +\frac{4x-1}{\Xi(x)^{3/2}}\right].
    \nonumber
\end{eqnarray}
The comparison between analytical and the numerical $B(E2)$
transition probabilities for a system of $N=40$ bosons, is shown
in Fig.~\ref{fig:be2}.
As for the excitation energies, there are divergences in the
$B(E2)$ values  at the critical point, although they now appear
even at the RPA order \cite{Rowe04_1}. However, at finite $N$
values no divergence should appear in the physical magnitudes or
their derivatives with respect to the control parameter $x$, even
at the critical point. This obvious remark allows us to determine
the nontrivial scaling exponents. Such an analysis was proposed in
Refs.~\cite{Dusuel04_3,Dusuel05_1,Dusuel05_2}, and we shall now
briefly recall how it works. The $1/N$ expansion of any physical
quantity $\Phi$ has two contributions, the regular (reg) and
singular (sing) respectively, when $x$ approaches the critical
value $x_\mathrm{c}=1/2$:
%
\begin{equation}
  \Phi_N(x)=
  \Phi_N^\mathrm{reg}(x)+\Phi_N^\mathrm{sing}(x).
\end{equation}
%
A close analysis of the singular part in the vicinity of the
critical point $x_\mathrm{c}$ shows that the singular part scales as
%
\begin{equation}
  \Phi_N^\mathrm{sing}(x)\simeq
  \frac{\Xi(x)^{\xi_\Phi}}{N^{n_\Phi}}
  \mathcal{F}_\Phi\left[N\Xi(x)^{3/2} \right],
\end{equation}
%
where $\mathcal{F}_\Phi$ is a function depending on the
scaling variable $N\Xi(x)^{3/2}$ only. To compensate the singularity
coming from $\Xi(x)^{\xi_\Phi}$ (or its derivative), one thus must
have $\mathcal{F}_\Phi (x) \sim x^{-2 \xi_\Phi /3}$ so that
$\Phi_N^\mathrm{sing}(x_\mathrm{c}) \sim
N^{-(n_\Phi+2\xi_\Phi/3)}$. In Table \ref{tab:exponents} the computed
scaling exponents for the low energy physical quantities studied are
summarized.
\begin{table}[t]
  \centering
  \begin{tabular}{|c|c|c|c|}
    \hline
    $\Phi$ & $\xi_\Phi$ & $n_\Phi$ & $-(n_\Phi+2\xi_\Phi/3)$\\
    \hline
    \hline
    $E_0$ & 1/2 & 0 & -1/3\\
    \hline
    $\Delta$ & 1/2 & 0 & -1/3\\
    \hline
    $\langle n_d \rangle$ & -1/2 & 0 & 1/3\\
    \hline
    $B(E2)$ & -1/2 & -1 & 4/3\\
    \hline
  \end{tabular}
  \caption{Scaling exponents for the ground state energy $E_0$, the gap  $\Delta$, the number of $d$-bosons in the ground-  state $\langle n_d  \rangle_{GS}$ and the $B(E2)$ transition probability.}
  \label{tab:exponents}
\end{table}

\begin{figure}[t]
  \centering
  \includegraphics[width=8cm]{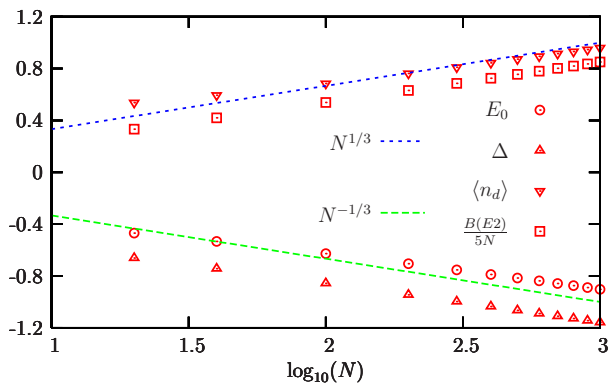}
  \caption{Plot of the singular parts of $E_0$, $\Delta$,
  $\langle n_d \rangle$ and $B(E2)/(5N)$ at the critical point $x_c=1/2$, in a
  $\log_{10}-\log_{10}$ scale. }
  \label{fig:scaling}
\end{figure}

In order to check these results, it is important to analyze the large $N$ behavior of $\Phi_N$. Therefore, we have numerically solved the problem by diagonalizing the boson Hamiltonian (\ref{eq:hamiltonian}) up to $N=1000$. Details of this calculation will be given in a forthcoming publication \cite{Dusuel05_4}. 
As can be seen in Fig.~\ref{fig:scaling}, an excellent agreement is found between the exponents predicted analytically and the numerical results. 
Let us underline that the scaling exponent for the ground state
energy has been recently obtained  by Rowe {\it et al.}
\cite{Rowe04_2} by using the collective model associated to the
IBM Hamiltonian \cite{Arias03}. This mapping onto a quartic
potential also explains why we found the same finite-size scaling
exponent for the ground state energy and the gap ($1/3$) in other
similar models \cite{Dusuel04_3,Dusuel05_1,Dusuel05_2}. However,
such an approach does not allow to simply compute the finite $N$
corrections and may not be suitable to obtain the behavior for
observables such as $B(E2)$. The CUTs method is thus, in this
context,  a very useful tool.


In the present work, we have exactly computed  finite-size
corrections beyond the RPA  in the symmetric phase of the IBM
model. We have shown that the spectral properties at the critical
point in the U(5)-O(6) transition have well-defined asymptotic
limits and we have calculated the $N$-dependent scale factors. 
A natural extension of this work would be to investigate the
broken phase ($x<1/2$) but the presence of Goldstone modes in the
low-energy spectrum (at the RPA level) makes it more involved \cite{Dusuel05_4}.

The study of the scaling properties at the critical point of a QPT of
a finite-$N$ particle model is of primer interest in several
mesoscopic systems like nuclei, molecules, and other physical
systems. The present results provide a tool to tackle such a study
and to characterize the approach of the system to the critical
regions as the number of particles goes to infinity.\\


\acknowledgments 
S. Dusuel gratefully acknowledges financial support of the DFG in SP1073. This work has been partially
supported by the Spanish DGI under projects number BFM2002-03315 , BFM2003-05316-C02-02, and  BFM2003-05316.



\end{document}